\documentclass[]{article}
\date{}
\usepackage{cite}
\pdfoutput=1
\usepackage{amsmath, amssymb, float}
\usepackage{graphicx}
\usepackage[total={6.5in, 8in}]{geometry}

\usepackage{abstract}

\usepackage{authblk}

\title{Symmetry Analysis of Surfactant Driven Thin Liquid Film Equations}

\author[1,*]{Devanshu Shekhar}
\author[2]{Satyananda Panda}

\affil[1]{\textit{Department of Physics, National Institute of Technology Calicut, Kerala, India}}
\affil[2]{\textit{Department of Mathematics, National Institute of Technology Calicut, Kerala, India}}
\affil[*]{\textit{Corresponding author: Devanshu Shekhar, devanshu2096@gmail.com}}

\begin{document}
	\twocolumn[
\maketitle
\centering
\begin{abstract}
	Spreading of liquid thin film driven by surfactant due to the Marangoni effect is
	described using a coupled
	system of second-order partial differential equations. Lie group of transformation
	are used to obtain the symmetries of the given system of partial differential equations.
	The symmetries are then used to arrive at a semi-analytic solution of the system. Furthermore, a vector field analysis of the obtained solution is performed to provide additional insights into the problem.
The obtained results demonstrate that the surfactant concentration drives the fluid, and thereby the fluid thins faster. 
	
\end{abstract}

\vspace{1cm}
]
\section{Introduction}
Differential equations arise everywhere in science and engineering.
The standard methods for solving differential equations analytically fall
short while dealing with equations describing the phenomenon of practical
interest, \textit{e.g.} problems stemming from industrial and medical applications.
A similar problem is of spreading of liquid thin film driven by an insoluble
surfactant due to \textit{marangoni effect} \cite{peterson2012simulation}. The
surfactant tends to decrease the surface tension on the surface of a fluid film
and creates a surface tension gradient. This results in a net
force per unit area, called \textit{Marangoni stress}, from a region of lower surface
tension to a region of higher surface tension. Marangoni effect has several
industrial and medical applications, \textit{e.g.},
\begin{itemize}
	\item In semiconductor industries, the drying of integrated circuits
	using alcohol is called \textit{Marangoni drying}. The water is expelled from a
	wet surface due to the Marangoni effect as the alcohol evaporates, leaving the area of contact dry.
	\item Pulmonary surfactants are present in a healthy lung to reduce the surface
	tension forces and prevent the lungs from collapse due to the Marangoni effect.
	Due to lack of surfactants, a patient experiences difficulty
	in breathing, which is called RDS (Respiratory Distress Syndrome) in medical terms.
	A remedy to this problem is SRT (Surfactant
	Replacement Therapy), which instills surfactants into the deficient lung to lower
	surface tension and improving wetting in the lung. 
\end{itemize}

The model spreading of surfactant dynamics on the solid substrate
ignoring surface tension has been
derived by Jensen and Grotberg~\cite{jensen1992insoluble}. Various numerical
methods have been developed to solve the model equations.
For example, Levy et al.~\cite{levy2006motion} used an implicit method
with Newton's iteration, and Mamoniat et al.~\cite{momoniat2013numerical} presented a
numerical scheme using an explicit
upwind scheme with Roe flux limiter. These  numerical methods
need to solve the  system at each time step with stability condition that imposes 
a smaller time step and require a long time for each solution.
We present in this work
a semi-analytic solution of the problem  with methods of
symmetry~\cite{arrigo2015symmetry, bluman2013symmetries, hydon2000symmetry},
that are quite useful for the easy computation
and interpretation of solution behaviour. 
The symmetry of a
differential equation  is a transformation that preserves
the form of the differential equation
in the new coordinate system but maps the solution
curves (or surfaces in case of partial differential equations) from one
coordinate system to another.  The symmetries form one-parameter Lie group of
transformation,\textit{ i.e.}, they depend upon a parameter $\epsilon $, which
assumes continuous values from a set $D$. In order to solve the equations, the
symmetries of the system are exploited to reduce the number of independent
variables by one. In our case the problem is reduced to a system of coupled
ordinary differential equations (ODEs), which are then easily solved using numerical methods.

The coupled partial differential equations (PDEs) describing liquid thin film and surfactant flow, such that the thin film height is given by $h = h(x, t)$ and the surfactant concentration is given by $\Gamma = \Gamma (x, t)$, in its simplest form \cite{momoniat2013numerical}, is given by
\begin{eqnarray} \label{pde}
\frac{\partial h}{\partial t}+\frac{\partial Q(x, t)}{\partial x}=0, \hspace{0.2cm} \frac{\partial \Gamma}{\partial t}+\frac{\partial P(x,t)}{\partial x}=0,
\end{eqnarray}
where, flux functions $P$ and $Q$ are given by,
\begin{equation}
Q(x,t)=\frac{h^3}{3}-\frac{h^2}{2} \frac{\partial \Gamma }{\partial x}, \hspace{0.2cm} P(x,t)=\frac{ \Gamma}{2} h^2-\Gamma h \frac{\partial \Gamma  }{\partial x}.
\end{equation}
In the next section, we present the method of symmetry analysis in brief. As the calculations are much involved, we have resorted to computer algebra for the analysis. In particular, the Sym package in Mathematica \cite{dimas2004sym} has been an invaluable aid throughout this work.
\section{Symmetry Analysis}
The Lie group of transformation,
\begin{equation}
\Phi _\epsilon : (x,y) \rightarrow (\hat{x},\hat{y})
\end{equation}
of a differential equation is an analytic function of the parameter $\epsilon $, such that, the\textit{ infinitesimal transformations} of the coordinates is given by
\begin{align}\label{tans}
\hat{t} &= t + \epsilon \xi ^1 (t, x, h, \Gamma ) + \mathcal{O}(\epsilon ^2)\nonumber \\
\hat{x} &= x + \epsilon \xi ^2 (t, x, h, \Gamma ) + \mathcal{O}(\epsilon ^2) \\
\hat{h} &= h + \epsilon \eta _1 (t, x, h, \Gamma ) + \mathcal{O}(\epsilon ^2)\nonumber\\
\hat{\Gamma } &= \Gamma + \epsilon \eta _2 (t, x, h, \Gamma ) + \mathcal{O}(\epsilon ^2)\nonumber,
\end{align}
where, $\{\xi _1, \xi _2, \eta_1, \eta_2\}$ are called \textit{infinitesimals} of the respective coordinates. The, Lie group of transformations are generated by an operator called \textit{infinitesimal generator}, given by,
\begin{equation}
X = \xi _1 \partial_t + \xi _2 \partial_x + \eta _1 \partial _h + \eta_2 \partial_\Gamma  .
\end{equation}
The partial derivatives determine the direction for the solution curve (or surface) in tangent space, just like $\hat{i}$ and $\hat{j}$ determine the direction of a point in coordinate space, and the infinitesimals, the coefficients, determine how much the curve moves in the respective direction under a symmetry transformation. In other words, the infinitesimal generators represent the tangent vector field in any coordinate system. Thus, a curve $F(x, y)$ is said to be invariant under symmetry transformations if and only if
\begin{equation}
	XF(x, y) = 0,
\end{equation}
and, in fact for a PDE, the solution we look for are these invariant curves (surfaces), because an invariant solution is still a solution! For the system in hand, the Lie group of transformation is generated by following set of infinitesimal generators:
\begin{multline} \label{symm}
\{X_1 = \partial _x, X_2 = \partial _t, X_3 = 2x\partial _x + 3 \Gamma \partial _{\Gamma } + h\partial _h, \\X_4 = 3t\partial _t - h\partial _h + x\partial _x\}.
\end{multline}
By the \textit{first fundamental theorem of Lie} \cite{bluman2013symmetries}, the Lie group of transformations, Eq.~\eqref{tans}, is equivalent to the solution of an initial value problem of a system of first order differential equations, 
\begin{equation} \label{lie}
\frac{d\hat{x}_i}{d\epsilon } = \xi _i(\hat{x}),
\end{equation}
with the initial condition,
\begin{equation}\label{ic}
\hat{x}_i(\epsilon = 0) = x_i.
\end{equation}
Here, $x_i \in \{x, t, h, \Gamma \}$ and $\xi _i \in \{ \xi _1, \xi _2, \eta_1, \eta_2\}$ respectively. $X_1$ and $X_2$ are trivial symmetries of the system corresponding to translation in space and time. Let's look at $X_3$. By Eq.~\eqref{lie}, the system of differential equations corresponding to Lie group of transformation is given by,
\begin{eqnarray}
\frac{d\hat{x}}{d\epsilon } = 2\hat{x}, \hspace{0.5cm} \frac{d\hat{\Gamma }}{d\epsilon } = 3\hat{\Gamma }, \hspace{0.5cm} \frac{d\hat{h}}{d\epsilon } =\hat{h},
\end{eqnarray}
and by applying the initial condition \eqref{ic}, we arrive at the following transformation,
\begin{eqnarray}
\hat{x} = e^{2\epsilon } x, \hspace{0.5cm} \hat{\Gamma } = e^{3\epsilon }\Gamma , \hspace{0.5cm} \hat{h} = e^{\epsilon }h,
\end{eqnarray}
\textit{i.e.}, the system has \textit{scaling symmetries}. Similar transformations are obtained from $X_4$. Now, in order to reduce the equations Eqs.~\eqref{pde} we can choose any linear combination of symmetries in Eq~\eqref{symm}, consequently, we have chosen symmetries such as to obtain most simplified reduced form. Choosing, $X = a X_3 + b X_1$, $a, b \in \mathbb{R}$ and applying \textit{invariant surface conditions},$$\left. X(\phi (x, t) - h) = 0\right |_{\phi (x, t) = h}$$ and $$\left. X(\psi (x, t) - \Gamma ) = 0\right |_{\psi (x, t) = \Gamma },$$ we arrive at the solutions,
\begin{equation} \label{h}
h(x,t) = f(t) \sqrt{2 a x+b},
\end{equation}
and,
\begin{equation}\label{g}
 \Gamma (x,t) =g(t) (2 a x+b)^{3/2}.
\end{equation}
The above solutions are then substituted into Eq.~\eqref{pde} to determine the functions $f(t)$  and $g(t)$ and we obtain,
\begin{equation} \label{ode1}
f'(t) = \frac{1}{2} a f(t)^2 (9 a g(t)-2 f(t)),
\end{equation}
and,
\begin{equation}\label{ode2}
g'(t) = \frac{5}{2} a f(t) g(t) (6 a g(t)-f(t)).
\end{equation}
Note that, the equations are independent of parameter $b$. Thus, the problem of solving a system of coupled partial differential equations of second order has been reduced to a problem of solving a system of coupled ordinary differential equations of first order. We have attempted to solve this system by one, converting it into a homogeneous equation and two, doing symmetry analysis. However, the level of difficulty we arrive at the end is same in both the case. For instance, for a particular case of $a = b = 1$,  dividing the Eqs.~\eqref{ode2} and \eqref{ode1}, we get, 
\begin{equation}
\frac{dg}{df} = \frac{30 f g^2 - 5 f^2 g}{9 f^2 g - 2 f^3},
\end{equation}
which is homogeneous.  Substituting $v = g/f$, we get,
\begin{equation}
f \frac{dv}{df} = \frac{21 v^2 - 3 v}{9 v - 2},
\end{equation}
which is not simple to solve. In addition, the solution obtained, after back substitution, is
\begin{equation} 
f = \frac{(g / f)^{3/2} c}{(1 - 7 (g / f))^{(5/21)}},
\end{equation}
where, c is constant of integration. It is not possible to separate $f$ and $g$ from the above equation. Consequently, it becomes impossible or at least extremely difficult to obtain an analytical solution for $f$ and $g$ in simple form. A similar situation arises while applying symmetry methods.
The numerical analysis, as it turned out, of the system gives satisfactory results without any complications. Such approach of reducing a complicated problem to a simpler one by symmetry methods and conducting numerical analysis on the reduced problem has been already made  \cite{momoniat2010symmetry}.
\section{Numerical Analysis}
\begin{figure}[H] 
	\centering
	\includegraphics[scale = 0.55]{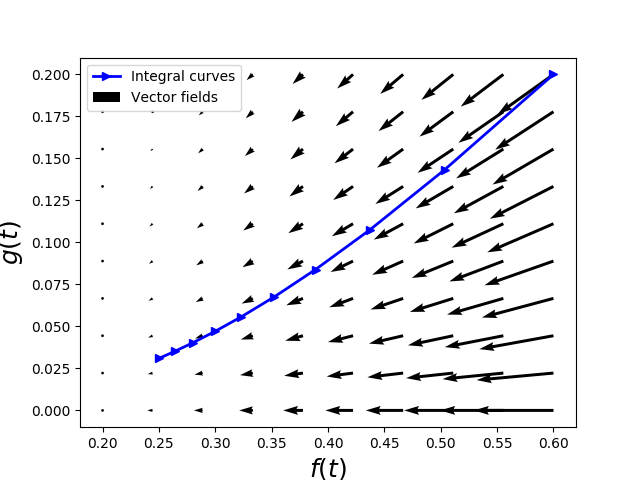}
	\caption{Phase-Space like representation of ODE system showing vector fields and an integral curve with initial condition Eqn.~\eqref{init}. }
	\label{fig:ph_space}
\end{figure}

Before we present the final solution of these system of equations, it is instructive to analyze the vector fields and integral curves of the system Eqs.~\eqref{ode1} and \eqref{ode2}. We consider functions $f(t)$ and $g(t)$ as components of vector $\textbf{f} = (f, g)$, such that, the system, reduces to
\begin{equation}
\frac{d\textbf{f}}{dt} = \Phi (\textbf{f}),
\end{equation}
where, $$\Phi (\textbf{f}) = (\frac{1}{2} a f^2 (9 a g-2 f), \frac{5}{2} a f g (6 a g-f))$$ is the tangent vector field in $f-g$ plane\footnote{In particular, while dealing with a second order ODE, it is usual to split the equation into two \textit{first} order ODE system. In that case the new variables are said to form the \textit{phase-space}.} and the solution to the system  Eqs.~\eqref{ode1} and \eqref{ode2} are the corresponding integral curves, as shown in Fig. \ref{fig:ph_space}.
The integral curve depends on the following initial conditions and the parameters $a$ and $b$, are chosen by a hit and trial method such that the solution is well behaved,
\begin{equation} \label{init}
f(0) = 0.6, \hspace{0.2cm} g(0) = 0.2, \hspace{0.2cm} a = 0.3, \hspace{0.2cm} b = 2.
\end{equation}
As can be seen, both the surfactant concentration and the free surface of the thin film diffuses and gradually goes to zero which is what is expected. 

\begin{figure}[H] 
	\centering
	\includegraphics[scale = 0.55]{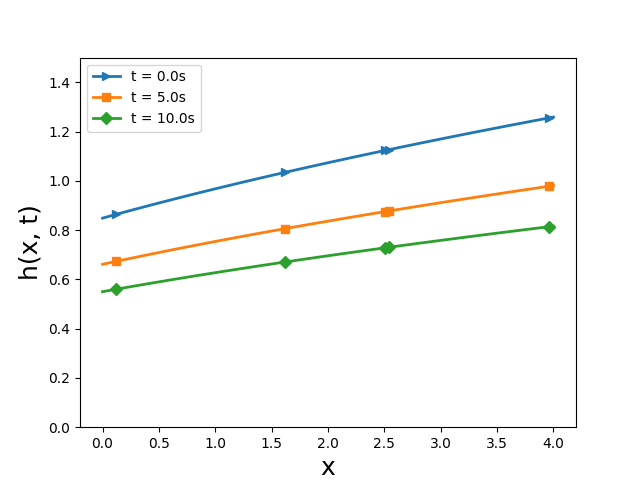}
	\caption{Evolution of thin-film height $h(x,t)$ at three different times. }
	\label{fig:im1}
\end{figure}

\begin{figure}[H] 
	\centering
	\includegraphics[scale = 0.55]{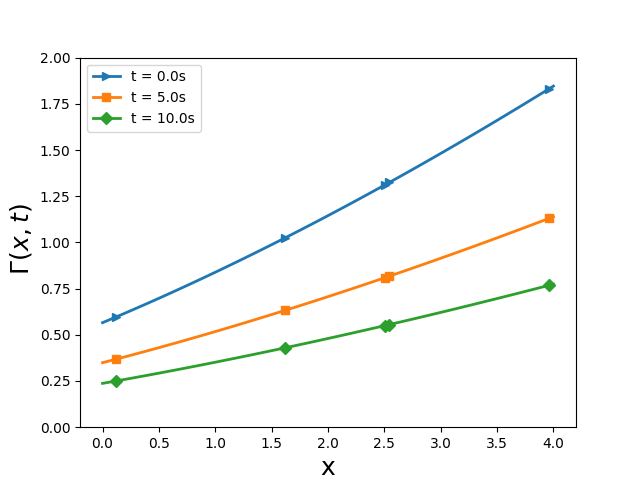}
	\caption{Evolution of surfactant concentration $\Gamma (x,t)$ at three different times. }
	\label{fig:im2}
\end{figure}

The numerical solutions are obtained in Mathematica using the \textit{NDSolve} function\cite{Mathematica}, with the initial conditions Eqn.~\eqref{init}. Substituting the numerical results in Eqs.~\eqref{h} and \eqref{g}, the final results obtained are shown in Figs.~\ref{fig:im1} and \ref{fig:im2}.
As can be observed, the increase in thin film height $h(x,t)$ and the surfactant concentration $\Gamma (x,t)$ is polynomial in $x$ and both diffuses with time as already pointed during the vector field analysis (Fig.~\ref{fig:ph_space}).
\section{Summary and Conclusions}
The Lie group of transformations describe the symmetries of a differential equation. The symmetries are basically transformations which map solution curve of a differential equation from one coordinate system to another. In particular, for partial differential equations, we look for surfaces which are invariant under such transformations. These are achieved by invoking the invariant surface conditions. These invariant solutions are then substituted back into the system to arrive at a system of ODEs, which are simply solved numerically.  In addition, the analysis of the ODEs using the concepts from vector calculus provided further insight into the problem. The semi-analytic solutions thus obtained are protected against instability and discontinuity, which are some of the problems encountered during numerical analysis.

In this paper, we have solved a coupled system of partial differential equations,
describing the flow of a  thin liquid  film driven by a surfactant due to the
Marangoni effect using the symmetry method. We have found that the obtained semi-analytic results predicted the real physical process where the fluid is driven by the surfactant and results in a decrease in the height of the free surface. Further, this work
shows that a mathematical tool can be easily obtained using the Lie group method for the
solution of partial differential equations representing the description of
similar problem in the fields of science and engineering. 
\bibliographystyle{plain}
\bibliography{research.bib}
\end{document}